\documentclass[pre,aps,twocolumn,superscriptaddress]{revtex4}
\usepackage{graphicx, amsmath, amsthm, amsfonts, amssymb, amscd, latexsym}

\begin{document}

\title{
Averaging of Nonlinearity Management with Dissipation}
\author{S. Beheshti, K. J. H. Law, P. G. Kevrekidis}
\affiliation{Department of Mathematics and Statistics, University of
Massachusetts, Amherst MA 01003-4515}
\author{Mason A. Porter}
\affiliation{Oxford Centre for Industrial and Applied Mathematics, Mathematical
Institute, University of Oxford, OX1 3LB, United Kingdom}

\begin{abstract}
Motivated by recent experiments in optics and atomic physics,
we derive an averaged nonlinear partial differential equation describing the dynamics of the complex field in a nonlinear Schr{\"o}dinger model in the presence of a periodic nonlinearity and a periodically-varying dissipation coefficient.  The incorporation of dissipation is motivated by experimental considerations.
We test the numerical behavior of the derived averaged equation by
comparing it to the original nonautonomous model in a prototypical case 
scenario and observe good agreement between the two.
\end{abstract}

\maketitle

\section{Introduction}
In the past few years, there has been an intense theoretical interest in the use of nonlinear Schr{\"o}dinger (NLS) equations to describe both the propagation of optical beams in waveguides
and fibers \cite{kivshar,reviewsopt} and the mean-field evolution of
Bose-Einstein condensates (BECs) \cite{book1,book2}.  Within this framework of dispersive equations that support solitary nonlinear waves, one of the particular topics of recent interest has been the
effects of spatially and/or temporally (i.e., in the evolution variable)
dependent nonlinearities. This subject, often called ``nonlinearity
management" \cite{borisbook} --- by analogy with the topic of ``dispersion management" that has been developed, in the context of optics, in far greater depth \cite{DMS} --- was originally proposed in the
study of layered optical media \cite{mezentsev}.  However, it has also
garnered considerable attention in the study of Bose-Einstein condensation,
where it was reformulated as Feshbach resonance management \cite{frm}.
 
Recent experimental work in optics has realized layered media through a concatenation of glass slides and air gaps.  This has allowed a more detailed examination of topics such as the breathing of localized pulses \cite{martin1} and the modulational instability of extended ones \cite{martin2,martin3}.  Moreover, in the context of Bose-Einstein condensates, the inter-atomic interactions (which are the source of
the nonlinearity at the mean-field level) can be adjusted experimentally in a very broad range
by employing either magnetic \cite{Koehler,feshbachNa} or optical Feshbach 
resonances \cite{ofr}. This has led to a significant number of both theoretical and experimental studies, including the formation (in the laboratory) of bright matter-wave solitons and soliton trains for
$^{7}$Li \cite{expb1,expb2} and $^{85}$Rb \cite{expb3} atoms.
On the theoretical side, such a modulation of the
interaction scattering length (and hence of the nonlinearity coefficient)
has been used, among other things, to stabilize attractive higher-dimensional BECs against collapse \cite{FRM1}.  More recently, spatial variations of the nonlinearity have also been considered.
In particular, it has been shown that such ``collisionally inhomogeneous'' condensates lead to a variety of interesting features, including adiabatic compression of matter-waves \cite{our1,fka}, 
Bloch oscillations of matter-wave solitons \cite{our1}, 
atomic soliton emission and atom lasers \cite{vpg12}, 
enhancement of transmittivity of matter-waves through barriers \cite{our2,fka2}, 
dynamical trapping of matter-wave solitons \cite{our2}, 
stable condensates exhibiting both attractive and repulsive interatomic 
interactions \cite{chin}, the delocalization transition of matter waves~\cite{LocDeloc}, and more.
Numerous different types of spatial variations of the nonlinearity have
now been considered, including linear \cite{our1,our2}, random \cite{vpg14}, 
periodic \cite{vpg16,LocDeloc,BludKon}, and localized (step-like)
\cite{vpg12,vpg17,vpg_new} ones.  There have also been a number of detailed
mathematical studies \cite{key-2,key-4,vprl} that address aspects
such as the effect of a ``nonlinear lattice potential" (i.e., a spatially 
periodic nonlinearity) on the stability/instability of solitary 
waves and the interplay between drift and diffraction/blow-up 
instabilities. 

In the case of fast variations of the nonlinearity coefficient as a function of the evolution variable (time in BECs and the propagation direction in optics), one successful strategy that has been employed is to average the nonautonomous, nonlinearity-managed dynamics to obtain an (averaged) autonomous system \cite{pelinovsky,pelinovsky1}. The stationary states \cite{pelinovsky2} and collapse 
properties \cite{pelinovsky3} of the latter can be analyzed in one- and in higher-dimensions, respectively. However, the presence of dissipation is a particularly important feature that 
arises when periodically varying the nonlinearity coefficients in both optics and BEC experiments and which was not incorporated in these earlier studies, 
to the best of our knowledge.  
In particular, in the optical case of nonlinearity
management there is a periodic loss (of the order of a few percent
of the intensity of the optical beam) every time the beam crosses
an interface between the different media (such as glass and air) \cite{martin1,martin2,martin3}. In BECs, if the Feshbach resonance is crossed, 
numerous atoms are lost, which 
results again in dissipative dynamics \cite{book1,book2}
(although it is important to note that it is not always necessary to cross
the actual resonance to change the sign of the scattering length, as vanishings of the scattering length
as a function of the external magnetic field can also occur away
from the resonance \cite{frm,FRM1}). 

Our goal in this Brief Report is to address the averaging of nonlinearly-managed dynamics in
the presence of dissipation.  The remainder of our presentation is organized as follows.
We first present the general setting in which both a periodic nonlinearity and a periodic
dissipation are applied. We then use averaging techniques to this
nonautonomous setting and obtain an autonomous partial differential equation (PDE)
describing the averaged dynamics.  We subsequently test the resulting
model against numerical experiments of the original dynamical equations,
obtaining good agreement between the two. Finally, we summarize our
findings and present some suggestions for possible future studies.

\section{Analytical Results}

Motivated by the above physical settings, we consider in our analysis
a time-dependent nonlinearity and include a time-dependent dissipation
term.  We thereby generalize the averaging technique of 
\cite{pelinovsky} and obtain a general averaged PDE in arbitrary dimensions.  
As our derivation hinges on the periodicity of the fast timescale, we require that 
the length of the period is at least an order of magnitude 
smaller than that of the slow scale over which we monitor the dynamics.

The primary model used for the physical settings we consider 
is an NLS equation of the form
\begin{equation}
	i u_t + \tfrac{1}{2}\Delta u + \gamma_0 |u|^2 u + i \zeta_0 u +
\tfrac{1}{\epsilon} \gamma \left( \tfrac{t}{\epsilon} \right) |u|^2 u
+ \tfrac{i}{\epsilon}\zeta \left( \tfrac{t}{\epsilon} \right)u = 0\,,   
\label{og}
\end{equation}
where $\textbf{x} \in \mathbb{R} ^n$, $t \in \mathbb{R}_+$, $\tau =
{t}/{\epsilon}$ and the quantities $\gamma_0$, $\zeta_0$ are parameters.  The continuous functions $\gamma (\tau)$, $\zeta (\tau)$ satisfy  
\begin{eqnarray*}
	\gamma(\tau+1)=\gamma(\tau), & \int_0^1\gamma(\tau)d\tau=0\,, \\
\zeta(\tau+1)=\zeta(\tau), & \int_0^1\zeta(\tau)d\tau=0\,.
\label{per}
\end{eqnarray*}
and represent the time-dependent part of the nonlinearity and dissipation,
respectively. The NLS equation describes the envelope of the electric field of
light in the context of optics, and it represents the mean-field wavefunction
of the BEC in the context of atomic physics.
To better understand the behavior of solutions, we use a multiple-scale expansion 
to derive an averaged equation for (\ref{og}) in arbitrary dimensions. 

Following the notation in Ref.~\cite{pelinovsky}, let $f_{-1}$ denote the zero-mean antiderivative of $f$.  It is given by
\begin{equation} \label{two}
	f_{-1}(\tau) = \int_0 ^\tau f(\tau ')d\tau ' - \int_0 ^1 \int_0 ^\tau f(\tau ') d\tau ' d\tau\,.
\end{equation}
Define the transformation
\begin{equation} \label{three}
	u(\textbf{x},t)={\rm e}^{-\zeta_{-1}(\tau)} {\rm e}^{(i j(\tau)
  |v(\textbf{x},t)|^2)} v(\textbf{x},t)\,, 
\end{equation}
where  $j(\tau) = \left(\gamma {\rm
  e}^{-2\zeta_{-1}(\tau)}\right)_{-1}$.  Using (\ref{two},\ref{three}), equation (\ref{og}) can be
  expressed as
\begin{eqnarray}\label{intNLS}
	iv_t-j(\tau)|v|_t ^2 v &=& -\tfrac{1}{2}\triangle v -\gamma_0 {\rm
  e}^{-2\zeta_{-1}(\tau)}|v|^2 v -  i\zeta_0 v \notag \\
&&  -\tfrac{1}{2} i j(\tau)
  \left[ 2 \left(\nabla |v|^2 \cdot \nabla v \right) + v \Delta |v|^2
  \right] \notag\\
 && +\tfrac{1}{2} j(\tau)^2 v \left( \nabla|v|^2 \cdot  \nabla|v|^2 \right)\,,
\end{eqnarray}
where $|v|_t ^2 = \frac{\partial}{\partial t}(|v|^2)$, $\Delta |v|^2$ stands for $\Delta (|v|^2)$, and $\nabla |v|^2$ stands for $\nabla (|v|^2)$.  

We isolate $|v|_t ^2$ by considering the expression $\overline{v}(\ref{intNLS}) - v \overline{(\ref{intNLS})}$ [i.e., Eq.~(\ref{intNLS}) times the complex conjugate of $v$ minus $v$ times the complex conjugate of Eq.~(\ref{intNLS})] to transform (\ref{og}) into the
standard form
\begin{eqnarray}\label{standardNLS}
	i v_t &=& -\tfrac{1}{2} \Delta v - \gamma_0 {\rm
 e}^{-2\zeta_{-1}(\tau)}|v|^2 v - i\zeta_0 v - 2 \zeta_0 j(\tau) |v|^2
 v \notag \\ 
 &&
-\tfrac{1}{2} i j(\tau) \left[ \Delta(|v|^2 v) - 2|v|^2 \Delta v + v^2
 \overline{\Delta v}  \, \right] \notag \\ 
 && - \tfrac{1}{2} j(\tau)^2 \left[ 2|v|^2\Delta |v|^2 + \nabla |v|^2 \cdot \nabla |v|^2 \right]v\,.
\end{eqnarray}
Using the multiple-scale expansion
$v(x,t) = w(x,t) + \epsilon v_1(x,t,\tau) + \mathcal{O}(\epsilon ^2)$, 
we obtain from (\ref{standardNLS}):
\begin{eqnarray}\label{zeroth}
	i w_t &=& - i v_{1\tau} -\tfrac{1}{2}\Delta w - \gamma_0 {\rm
 e}^{-2\zeta_{-1}(\tau)}|w|^2 w \notag \\
&& - i\zeta_0 w - 2 \zeta_0 j(\tau) |w|^2 w \notag \\ 
 && -\tfrac{1}{2} i j(\tau) \left[ \Delta(|w|^2 w) - 2|w|^2 \Delta w + w^2
 \overline{\Delta w}  \, \right] \notag \\ 
 && - \tfrac{1}{2} j(\tau)^2 \left[ 2|w|^2\Delta |w|^2 + \nabla |w|^2 \cdot \nabla |w|^2 \right]w\,.
\end{eqnarray}
Integrating (\ref{zeroth}) yields an expression that one can reintroduce
  into the equation to solve for $v_{1\tau}$.  Consequently, the averaged
equation for (\ref{og}) takes the form 
\begin{eqnarray}\label{ave}
	i w_t &=& - \tfrac{1}{2}\Delta w - \gamma_0 \rho |w|^2 w - i \zeta_0 w 
\notag \\
	&& - \tfrac{\sigma^2}{2} w \left[\nabla |w|^2 \cdot \nabla |w|^2 + 2|w|^2\Delta |w|^2 \right]\,,
\end{eqnarray}
with $\sigma^2=\int_0^1 j(\tau)^2 d\tau$ and $\rho = \int_0^1 e^{-2
  \zeta_{-1}(\tau)} d \tau$.  Observe that the formal expansion
$v=w+\epsilon v_1$ yields an equation which no longer depends upon the
  fast time-scale $\tau$ (that is, an autonomous PDE). 
This approach also enables us to obtain the governing 
dynamics for the leading-order correction to the averaged behavior:
\begin{eqnarray}\label{correction}
	v_1 &=& -\tfrac{1}{2}\left[ \Delta(|w|^2 w) - 2|w|^2 \Delta w + w^2
 \overline{\Delta w} \, \right] j_{-1}(\tau)\notag \\
&&  + \tfrac{1}{2}i w \left[ \nabla |w|^2 \cdot \nabla |w|^2 +
 2|w|^2\Delta |w|^2  \right]\left(j(\tau)^2-\sigma ^2 \right)_{-1}
 \notag \\
&& + i \gamma_0 |w|^2 w \left( {\rm e}^{-2\zeta_{-1}(\tau)}
 \right)_{-1}+2i\zeta_0  j_{-1}(\tau) |w|^2 w\,,
\end{eqnarray}
which can be compared with Eq.~(2.13) in Ref.~\cite{pelinovsky}.

\section{Numerical Corroboration}

In order to test the validity of Eq.~(\ref{ave}) for the averaged dynamics, we implement 
a prototypical two-dimensional realization of the above setting
with radial symmetry (following the lines of \cite{FRM1}).  
In particular, we consider the case of the two-dimensional 
unstable (against collapse) NLS soliton --- the so-called Townes soliton 
\cite{townes}) --- with focusing nonlinearity.  It has been shown in the context
of BECs that such a solution can 
be stabilized using a rapidly oscillating nonlinearity coefficient \cite{FRM1}.  
The resulting dynamics contain a fast time-scale periodicity associated with 
the nonlinearity management, so that this setting provides an ideal testbed 
for examining the accuracy of Eq.~(\ref{ave}).

\begin{figure}
\begin{center}
\includegraphics[width=80mm]{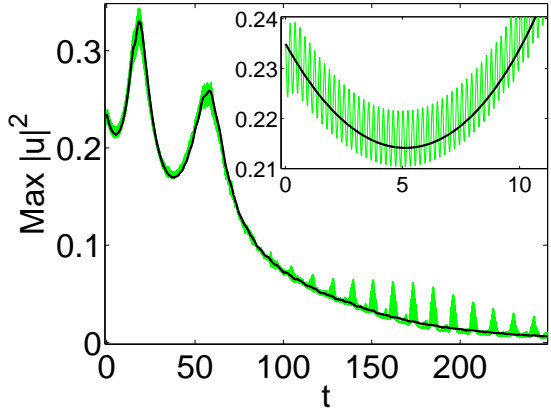}
\end{center}
\caption{(Color online) Averaged dynamics of Eq. (\ref{ave}) versus 
full dynamics of Eq. (\ref{num1}):  the maximum of the intensity of (\ref{num1}) is shown in gray (green in the online version).  We use the parameters $a_0=2\pi$, $a_1=-8\pi$, $\omega=30$, $b=b_0\times10^{-4}$, and $b_0=2$.  
Overlayed (in black) is the same diagnostic for the averaged equation.  It is clear that the averaged dynamical behavior is captured accurately over 250 time units. The inset shows a magnification of the small time-scale oscillations that are responsible for the stabilization and the overlayed average curve over the first 12 time units. }
\label{kfig1}
\end{figure}

As in Refs.~\cite{FRM1}, we consider the setting
\begin{equation}
	\left[i\partial_t+\frac{1}{2}\Delta+\gamma(t)|u|^2+i\zeta(t)\right]u=0\,,  
\label{num1}
\end{equation}
where the time-dependent nonlinearity $\gamma(t)$ and the time-dependent
dissipation $\zeta(t)$ (the latter was absent in \cite{FRM1}) are
given by
\begin{eqnarray}
	\gamma(t) &=& a_0+a_1 \sin(\omega t) \notag \\
\zeta(t) &=& b[1-\cos(\omega t)]\,.
\label{num2}
\end{eqnarray}

\begin{figure}
\begin{center}
\includegraphics[width=80mm]{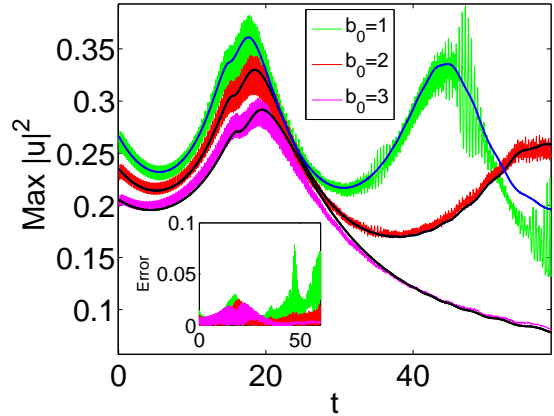}
\end{center}
\caption{(Color online) Three comparisons like that of Fig.~\ref{kfig1} over the first 60 time units with $b_0=1, 2$, and $3$ for the top, middle, and bottom curves, respectively.  All other parameters are the same as in Fig.~\ref{kfig1}.  In particular, observe in the top comparison that the averaged equation (\ref{ave}) begins diverging from the long time-scale dynamics (in the case of the smaller-magnitude dissipation term).  The inset shows the absolute error between the solution of the full equation, $u_{\rm full}$, and that of the averaged, $u_{\rm ave}$, Error$=|{\rm max}(|u_{\rm ave}|^2)-{\rm max}(|u_{\rm full}|^2)|(t)$. One can see more clearly here that after $t=30$ the error in the case with the smallest dissipation coefficient grows considerably.}  
\label{kfig2}
\end{figure}

Let $\epsilon=2\pi/\omega$, $f_0=\frac{1}{\epsilon}\int_0^\epsilon f(t) dt$, and $\tilde{f}(\tau)=\epsilon[f(\tau)-f_0]$, where $\tau$ is defined below Eq.~(\ref{og}), for any function $f$.  Applying this operation to the functions $\gamma(\epsilon \tau)$ and $\zeta(\epsilon \tau)$ brings Eq.~(\ref{num1}) in
the form of Eq.~(\ref{og}) (upon subsequently dropping the tildes) and allows us to apply Eq.~(\ref{ave}) to this setting.  
$\zeta(t)>\zeta_{\rm crit} \approx 5.8$ (in the regime of instability) \cite{FRM1,adhikari} for more than half the period.  Figure \ref{kfig1} shows the maximum intensity $|u|^2$ of the field in the full equation (\ref{num1}) in gray
(green in the color online version) and the intensity $|w|^2$ of the averaged equation (\ref{ave}) in black.  It clearly illustrates the strong correlation between the latter and the average (over a fast period) of the
former.  In Fig.~\ref{kfig2}, we illustrate this agreement over the initial few periods of the
macroscopic dynamics of the equation for several different values of
$b=b_0 \times 10^{-4}$, with $b_0=1,2,3$. It is clear that as the
magnitude of the dissipation is increased, the amplitude of the
nonlinear solution sustains a stronger decrease and a weaker 
oscillatory behavior. Nevertheless, the agreement between the average
of the full dynamics (over the fast scale) and the proposed averaged dynamics remains essentially satisfactory in all the studied cases.

\section{Conclusions}

In conclusion, we have considered the physically
realistic case of periodic dissipative dynamics in the setting of 
periodically-managed nonlinearity. We have argued on physical grounds
that the inclusion of dissipation is relevant for both optically
layered media and for Bose-Einstein condensates crossing Feshbach resonances. We have systematically generalized earlier works by obtaining a PDE that incorporates both the average
and the fluctuating parts of the dissipation.  We showed that this generalized PDE model, which is valid in both one-dimensional and multi-dimensional settings, is in good agreement with the average of the original dynamical equation for different dissipation characteristics within a prototypical 
two-dimensional nonlinearity management setting proposed earlier.

As discussed in the introduction, in addition to the case of temporally-dependent nonlinearity explicitly considered here, there has been a lot of recent attention on spatially-dependent microstructures in the nonlinearity. It would thus be interesting to extend our considerations to the case of spatially-dependent nonlinearity prefactors and to derive the corresponding ``averaged'' dynamics (provided that 
the nonlinearity prefactor varies over a fast spatial scale). Extending
such averaging considerations to spatially and spatio-temporally varying
nonlinearities is an interesting endeavor under current consideration, and relevant results will be reported in future studies.

{\it Acknowledgements}. 
PGK gratefully acknowledges support from NSF-CAREER, NSF-DMS-0505663, 
NSF-DMS-0619492 and the Alexander von Humboldt Foundation.

\end{document}